\documentclass[a4paper, 11pt]{article}
\usepackage[T1]{fontenc}
\usepackage{graphicx,epsf,amssymb,amsbsy,amsfonts,amssymb,amsmath,physics,courier,IEEEtrantools,jheppub,romannum,ytableau,empheq}

\def\be{\begin{equation}}
\def\ee{\end{equation}}
\def\bea{\begin{eqnarray}}
\def\eea{\end{eqnarray}}

\def\l{\lambda}
\def\s{\sigma}\def\l{\lambda}

\def\bg{\bar{g}}
\def\beq{\begin{eqnarray}}\def\eeq{\end{eqnarray}}
\def\ba#1\ea{\begin{align}#1\end{align}}
\def\bg#1\eg{\begin{gather}#1\end{gather}}
\def\bm#1\em{\begin{multline}#1\end{multline}}
\def\bmd#1\emd{\begin{multlined}#1\end{multlined}}

\def\l{\lambda}

\def\s{\sigma}

\def\nn{\nonumber}

\def\({\left(}
\def\){\right)}
\def\[{\left[}
\def\]{\right]}

\def\l{\lambda}

\def\s{\sigma} 
\def\om{\omega}

\pdfoutput=1

\title{Modified celestial amplitude in Einstein gravity}
\author[1]{Shamik Banerjee,} 
\author[2]{Sudip Ghosh,} 
\author[1,4]{Pranjal Pandey,} 
\author[3,4]{Arnab Priya Saha}
\affiliation[1] {Institute of Physics, \\ Sachivalaya Marg, Bhubaneshwar, India-751005}
\affiliation[2]{Okinawa Institute of Science and Technology, 1919-1 Tancha, Onna-son, Okinawa 904-0495, Japan}
\affiliation[3]{Harish-Chandra Research Institute, \\ Chhatnag Road, Jhunsi, Allahabad, India-211019}
\affiliation[4]{Homi Bhabha National Institute, Anushakti Nagar, Mumbai, India-400085}
\emailAdd{banerjeeshamik.phy@gmail.com, pranofmpvm@gmail.com, arnabpriyasaha@hri.res.in, \\ sudip112phys@gmail.com}

\abstract {In this paper we evaluate the modified celestial amplitude for gravitons and gluons, as defined in arXiv:1801.10171[hep-th]. We find that the modified (tree) amplitude is finite for gravitons in Einstein gravity. The modified amplitude behaves like correlation function of operators inserted at various points of null-infinity in the Minkowski space-time. Therefore, unlike the standard celestial amplitudes, these are three dimensional objects. We also show that this amplitude admits conformal soft factorization recently studied in the literature.}

\begin{document}
\maketitle
\flushbottom
\section{Introduction}

In flat space holography the natural observables are the $S$-matrix elements. They are the analogues of the correlation functions of the boundary CFT in the AdS/CFT correspondence. In a holographic description it is expected that the global symmetries of the dual theory should match the asymptotic symmetries of the bulk theory of quantum gravity. When the bulk is four dimensional asymptotically flat space-time, the asymptotic symmetry group is the extended BMS group \cite{Bondi:1962px,Sachs:1962zza,Strominger:2013jfa,Barnich:2009se,Kapec:2014opa,Campiglia:2015kxa} which, besides supertranslation, also contains superrotation. Superrotations are local conformal transformations acting on the two dimensional celestial sphere.  This is an extension of the Lorentz group $SL(2,\mathbb{C})$ which acts on the celestial sphere as the group of \textit{global} conformal transformations. Therefore, from a holographic perspective, it is desirable to have a (complete) set of observables which transform naturally under the (Lorentz) conformal group. In order to achieve this \cite{Pasterski:2016qvg,Pasterski:2017kqt,Pasterski:2017ylz} has put forward a very interesting proposal in which, instead of plane-waves, one uses the conformal primary wave-functions to describe the states of the incoming and outgoing particles in an $S$-matrix element. To be more precise, for massless particles, the change of basis is given by \cite{Pasterski:2016qvg,Pasterski:2017kqt,Pasterski:2017ylz},
\be
\tilde S\big(\{z_i, \bar z_i, \lambda_i, \sigma_i\}\big) = \prod_{i=1}^{n} \int_{0}^{\infty} d\omega_i \ \omega_i^{i\lambda_i} S\big(\{\omega_i,z_i,\bar z_i, \sigma_i\}\big)
\ee
Here $\sigma_i$ denotes the helicity of the $i$-th particle \footnote{In a later part we will denote the simplex variables \cite{Pasterski:2017ylz} also by $\sigma$. This should be clear from the context.} and the on-shell momenta are parametrized as,
\be\label{para}
p_i = \omega_i (1+z_i\bar z_i, z_i + \bar z_i , -i(z_i - \bar z_i), 1- z_i \bar z_i), \quad p_i^2 = 0
\ee

Under $SL(2,\mathbb C)$ (Lorentz) transformation $\omega$ and $z$ transforms as,
\be
\omega \rightarrow \omega |cz+d|^2 , \quad z \rightarrow \frac{az+b}{cz+d}, \quad \bar z \rightarrow c.c,  \qquad 
\begin{pmatrix}
a & b \\
c & d
\end{pmatrix} \in SL(2,\mathbb{C})
\ee

The new amplitude $\tilde S\big(\{z_i, \bar z_i, \lambda_i, \sigma_i\}\big)$ is a Mellin transformation of the standard $S$-matrix element $S\big(\{\omega_i,z_i,\bar z_i, \sigma_i\}\big)$ and is known as the celestial amplitude because $(z_i,\bar z_i)$ can be thought of as complex coordinates of points on the celestial sphere. Under Lorentz transformation $\tilde S$ transforms covariantly,
\be\label{qp}
\tilde S\big(\{z_i, \bar z_i, \lambda_i, \sigma_i\}\big) = \prod_{i=1}^{n} \frac{1}{(cz_i + d)^{2h_i}} \frac{1}{(\bar c \bar z_i + \bar d)^{2\bar h_i}} \tilde S\bigg(\frac{az_i+b}{cz_i+d} \ ,\frac{\bar a \bar z_i + \bar b}{\bar c \bar z_i + \bar d} \ ,\lambda_i \ ,\sigma_i\bigg)
\ee
where,
\be
h=\frac{1+i\lambda - \sigma}{2} , \quad \bar h = \frac{1+i\lambda +\sigma}{2}
\ee
Equation \eqref{qp} shows that $\tilde S$ transforms like the correlation function of conformal (quasi) primaries of weights $(h_i,\bar h_i)$, inserted at the points $(z_i,\bar z_i)$ on the celestial sphere. This transformation law is natural because the Lorentz group acts on the celestial sphere as the group of global conformal transformations. The action of global space-time translations on the Mellin-amplitude $\tilde S$ was studied in \cite{Stieberger:2018onx}.

Now, a modified form of Mellin amplitude was defined in \cite{Banerjee:2018gce,Banerjee:2018fgd},
\be\label{modmellin}
\boxed{
\mathcal A\big(\{u_i,z_i, \bar z_i, \lambda_i, \sigma_i\}\big) = \prod_{i=1}^{n} \int_{0}^{\infty} d\omega_i \ \omega_i^{i\lambda_i} e^{-i\varepsilon_i\omega_i u_i}S\big(\{\omega_i,z_i,\bar z_i, \sigma_i\}\big)}
\ee
where $\varepsilon_i = +1$ for an outgoing particle and $\varepsilon_i = -1$ for an incoming particle. The $u$ coordinate has the interpretation of (retarded) time and $(u,z,\bar z)$ can be thought of as coordinates at null-infinity in asymptotically flat space. This interpretation is further confirmed by the transformation property of $(u,z,\bar z)$ under Lorentz transformation \cite{Banerjee:2018gce,Banerjee:2018fgd},
\be
u \rightarrow \frac{u}{|cz+d|^2} , \quad z \rightarrow \frac{az+b}{cz+d} , \quad \bar z \rightarrow c.c
\ee

The modified Mellin transformation \eqref{modmellin} can also be inverted in the standard way to recover the standard $S$-matrix element,
 \be\label{inv}
\boxed{
S\big(\{\omega_i,z_i,\bar z_i, \sigma_i\}\big) = e^{i\sum_{i=1}^{n}\varepsilon_i\omega_i u_i} \prod_{i=1}^n \int_{-\infty}^{\infty} \frac{d\lambda_i}{2\pi} \ \omega_i^{-i\lambda_i -1} \mathcal A(\{u_i,z_i, \bar z_i, \lambda_i, \sigma_i\})}
\ee 

Under (Lorentz) conformal transformation the Mellin amplitude $\mathcal A$ transforms as,
\be
\mathcal A\big(\{u_i,z_i, \bar z_i, \lambda_i, \sigma_i\}\big) = \prod_{i=1}^{n} \frac{1}{(cz_i + d)^{2h_i}} \frac{1}{(\bar c \bar z_i + \bar d)^{2\bar h_i}} \mathcal A\bigg(\frac{u_i}{|cz_i + d|^2} \ , \frac{az_i+b}{cz_i+d} \ ,\frac{\bar a \bar z_i + \bar b}{\bar c \bar z_i + \bar d} \ ,\lambda_i \ ,\sigma_i\bigg)
\ee
and under global space-time translation,
\be
\mathcal A\big(\{u_i + f(z_i,\bar z_i),z_i, \bar z_i, \lambda_i, \sigma_i\}\big) = \mathcal A\big(\{u_i,z_i, \bar z_i, \lambda_i, \sigma_i\}\big)
\ee
where
\be
f(z,\bar z) = a + bz + \bar b \bar z + c z\bar z
\ee

For the sake of completeness, let us now briefly explain the origin of the modified transformation equation \eqref{modmellin}. In the usual $S$-matrix the asymptotic states are described by the direct product of the single particle Wigner states $\ket{p,\sigma}$ where $p$ is an on-shell momentum and $\sigma$ is the helicity. Here $p$ is parametrized as in \eqref{para}. Now, in the Mellin amplitude $\tilde S\big(\{z_i, \bar z_i, \lambda_i, \sigma_i\}\big)$ the asymptotic states are described as the direct product of the single particle states $\ket{z,\bar z,\lambda,\sigma}$ which are the Mellin transform of the $\ket{p,\sigma}$ states \cite{Pasterski:2016qvg,Pasterski:2017kqt,Pasterski:2017ylz,Banerjee:2018gce}, i.e,
\be
\ket{z,\bar z, \lambda, \sigma} = N \int_0^{\infty} d\omega \omega^{i\lambda} \ket{p(\omega,z,\bar z),\sigma}
\ee 
where $N$ is a normalization constant. Now, given these states, one can calculate the quantum mechanical transition amplitude given by \cite{Banerjee:2018gce} ,
\begin{equation}\label{amp}
\begin{aligned}
\bra{z,\bar z,\lambda,\sigma} e^{-iH(U-U')} \ket{z',\bar z',\lambda',\sigma'} 
&= \bra{u,z,\bar z,\lambda,\sigma}\ket{u',z',\bar z',\lambda',\sigma'} \\
&= \lim_{\delta\rightarrow 0+} \frac{\delta_{\sigma \sigma'}}{2\pi} \frac{\Gamma\big(i(\lambda' - \lambda)\big)\delta^2(z' -z)}{\big(-i(u'-u + i\delta)\big)^{i(\lambda' - \lambda)}}
\end{aligned}
\end{equation}
where $u = (1+z\bar z) U$ and we have also defined the Heisenberg picture basis state, 
\be
\ket{u,z,\bar z,\lambda,\sigma} = e^{iHU}\ket{z,\bar z,\lambda,\sigma} = N \int_0^{\infty} d\omega \omega^{i\lambda} e^{i\omega u} \ket{p(\omega,z,\bar z),\sigma}
\ee
The action of the Poincare group on the states $\ket{u,z,\bar z,\lambda,\sigma}$ is simply given by,
\be
U(\Lambda) \ket{u,z,\bar z,\lambda,\sigma} = \frac{1}{(cz+d)^{2h}}\frac{1}{(\bar c \bar z + \bar d)^{2\bar h}} \ket{\frac{u}{|cz+d|^2}, \frac{az+b}{cz+d},\frac{\bar a \bar z + \bar b}{\bar c \bar z + \bar d},\lambda,\sigma}
\ee
and
\be
e^{-iP.a} \ket{u,z,\bar z,\lambda,\sigma} = \ket{u+f(z,\bar z),z,\bar z}
\ee
where $f(z,\bar z) = (a^0-a^3) - (a^1 - i a^2)z - (a^1+ia^2) \bar z + (a^0 + a^3) z\bar z$. 

This suggests a picture in which the particle is moving in a three dimensional space-time with coordinates $(u,z,\bar z)$ on which the Poincare group acts geometrically, i.e, 
\be
(u,z,\bar z) \xrightarrow{L.T} \bigg(\frac{u}{|cz+d|^2}, \frac{az+b}{cz+d},\frac{\bar a \bar z + \bar b}{\bar c \bar z + \bar d}\bigg)
\ee
\be
(u,z,\bar z) \xrightarrow{Translation} (u+ (a^0-a^3) - (a^1 - i a^2)z - (a^1+ia^2) \bar z + (a^0 + a^3) z\bar z , z, \bar z )
\ee
In fact, this is exactly the way in which the Poincare group acts on null infinity in Minkowski space parametrized by the Bondi coordinates $(u,z,\bar z)$. One can also check that as expected, the transition amplitude \eqref{amp} is (covariant) invariant under the Poincare transformation. 

Now, \emph{the modified Mellin amplitude $\mathcal A\big(\{u_i,z_i, \bar z_i, \lambda_i, \sigma_i\}\big)$ is essentially the $S$-matrix element when the asymptotic in and out states are described by free particles moving in this three dimensional space with coordinates $(u,z,\bar z)$}. To be more precise, the states of the incoming or outgoing particles are now given by the direct product of the states,
\be
\ket{u,z,\bar z,\lambda,\sigma, in/out} = N \int_0^{\infty} d\omega \omega^{i\lambda} e^{i\omega u} \ket{p(\omega,z,\bar z),\sigma,in/out}
\ee
The main difference between the Mellin transforms $\tilde S$ and $\mathcal A$ is that in $\tilde S$ all the incoming and outgoing particles are constrained to lie on the same equal-$u$ hyper surface, whereas in $\mathcal A$ the particles are separated in time. Let us now explain the difference between $\tilde S$ and $\mathcal A$ from the point of view of the (Minkowski space) wave functions of the asymptotic particles. For simplicity we consider the external particles to be scalars. 

Now, as we have already discussed, the Mellin amplitude $\tilde S$ describes the scattering process when the asymptotic particles are described by the conformal primary wave functions \cite{Pasterski:2016qvg,Pasterski:2017kqt,Pasterski:2017ylz}, 
\be\label{PSS}
\Phi_{\Delta}^{\pm} (x^{\mu} | z, \bar z) = \frac{(\mp i)^{\Delta} \Gamma(\Delta)}{(- q(z,\bar z) \cdot x \mp i\delta)^{\Delta}} \ ,  \quad  \Delta = 1 + i\lambda, \quad \delta\rightarrow 0+
\ee
where $q^{\mu}(z,\bar z) = (1+z\bar z, z + \bar z , -i(z - \bar z), 1- z \bar z)$ is a unit null-vector. The superscript $\pm$ refers to the Mellin transformation of positive and negative energy plane waves, respectively. Now, the wave function \eqref{PSS} is distinguished by the fact that it is singular along the null-hyperplane, $x\cdot q(z,\bar z)=0$, which passes through the origin of the Minkowski space. In other words, one can say that the particle described by \eqref{PSS} is localised on the null-hyperplane, $x\cdot q(z,\bar z)=0$. Now, since the Klein-Gordon equation is Poincare invariant, it is natural to consider other solutions  in which the particle is localised on different null-hyperplanes in Minkowski space. The complete set of such solutions can be parametrized as \cite{Banerjee:2018gce}, 
\be\label{ba}
\Phi_{\Delta}^{\pm} (x^{\mu} |u, z, \bar z) = \frac{(\mp i)^{\Delta} \Gamma(\Delta)}{(- q(z,\bar z)\cdot x + u \mp i\delta)^{\Delta}}, \quad  -\infty < u < \infty, \quad \delta \rightarrow 0+
\ee
The particle described by the wave function \eqref{ba} is now localized along the null-hyperplane, $x\cdot q(z,\bar z) - u = 0$ and by varying $(u,z,\bar z)$ we can generate all the null-hyperplanes in the Minkowski space. From the Poincare transformation property of the wave function \eqref{ba} one can conclude that the three parameters $(u,z,\bar z)$ have the interpretation of Bondi coordinates at null-infinity \cite{Banerjee:2018gce}. The modified Mellin amplitude $\mathcal A$ describes the scattering process when the asymptotic particles are described by wave functions \eqref{ba}. To be more precise, for the $\tilde S$ amplitude the wave functions are $\big\{\Phi_{\Delta_i}^{\pm} (x^{\mu} |u, z_i, \bar z_i)\big\}$. Since the retarded time $u$ is the \emph{same for all the particles}, we can set $u=0$ by time translation invariance. On the other hand, the $\mathcal A$ amplitude is computed with the external particle wave functions $\big\{\Phi_{\Delta_i}^{\pm} (x^{\mu} |u_i, z_i, \bar z_i)\big\}$ and so \emph{different particles are now localized at different retarded times}. The \emph{separation in retarded time is the main difference between the $\mathcal A$ and $\tilde S$ amplitudes}.

The rest of the paper is organized as follows. In section-\eqref{reg} we discuss the role of $i\delta$ in regularizing the modified amplitude $\mathcal A$. In section-\eqref{MellinAmps} we evaluate $\mathcal{A}$ corresponding to $4$ - graviton and $4$ - gluon MHV amplitudes. We find that for gravitons $\mathcal A$ is finite even in Einstein gravity. This is a welcome feature from a holographic perspective. We then go on to study the conformal soft factorization property of $\mathcal A$. Soft factorization is non-trivial in the Mellin basis because one integrates over the energies of the asymptotic particles. It has been proposed and studied in \cite{Fan:2019emx,Nandan:2019jas,Pate:2019mfs,Donnay:2018neh,Adamo:2019ipt,Puhm:2019zbl,Guevara:2019ypd} that soft limit in the Mellin amplitude corresponds to taking $\lambda_p \rightarrow 0$ if the $p$-th particle is going soft. To be more precise, it has been shown that the limiting value of $i\lambda_p S\big(\{z_i, \bar z_i, \lambda_i, \sigma_i\}\big)$, as $\lambda_p\rightarrow 0$, factorizes. We show in section-\eqref{csf} that analogous (conformal) soft factorization is also true for $i\lambda_p \mathcal A\big(\{u_i,z_i, \bar z_i, \lambda_i, \sigma_i\}\big)$, as $\lambda_p\rightarrow 0$. 

\section{Regularization}\label{reg}
The modified Mellin amplitude $\mathcal A$ has the form,
\be
\mathcal A\big(\{u_p,z_p, \bar z_p, \lambda_p, \sigma_p\}\big) = \prod_{p=1}^{n} \int_{0}^{\infty} d\omega_p \ \omega_p^{i\lambda_p} e^{-i\varepsilon_p\omega_p u_p}S\big(\{\omega_p,z_p,\bar z_p, \sigma_p\}\big)
\ee
For large values of energy the exponential factors $e^{\pm i\omega u}$ oscillate rapidly. These can be regulated by adding a small imaginary part to $u$. To be more precise, we make the following shift,
\be
u_p \rightarrow u_p - i \varepsilon_p \delta, \qquad \delta \rightarrow 0+  
\ee
The origin of the convergence factor can also be traced back to the conformal primary wave-functions \eqref{ba},
\begin{equation}
\Phi_{\Delta}^{\pm} (x^{\mu} |u, z, \bar z)  = \frac{(\mp i)^{\Delta} \Gamma(\Delta)}{(- q(z,\bar z)\cdot x + u \mp i\delta)^{\Delta}} 
\end{equation}
We can see that effectively $u$ has a small imaginary part $\propto\delta$, whose sign depends on the sign of the energy. 
  
Therefore the amplitude $\mathcal A$ should be understood as, 
\be
\begin{gathered}
\mathcal A\big(\{u_p,z_p, \bar z_p, \lambda_p, \sigma_p\}\big) = \prod_{p=1}^{n} \int_{0}^{\infty} d\omega_p \ \omega_p^{i\lambda_p} e^{-i\varepsilon_p\omega_p (u_p - i\varepsilon_p\delta)} S\big(\{\omega_p,z_p,\bar z_p, \sigma_p\}\big) \\
= \prod_{p=1}^{n} \int_{0}^{\infty} d\omega_p \ \omega_p^{i\lambda_p} e^{-i\varepsilon_p\omega_p u_p} e^{-\delta s} S\big(\{\omega_p,z_p,\bar z_p, \sigma_p\}\big), \quad \delta\rightarrow 0+
\end{gathered}
\ee
where
\be
s = \sum_p \omega_p > 0
\ee
This provides the necessary damping for the integrals to be well defined and plays a crucial role in the rest of the paper. In most of the formulas below, we will not explicitly write down the convergence factor but it should always be kept in mind.



\section{Modified Mellin Amplitudes}\label{MellinAmps}
Mellin transform of scattering amplitudes in the absence of the retarded time has been studied in detail in \cite{Pasterski:2016qvg,Pasterski:2017ylz,Schreiber:2017jsr,Cardona:2017keg,Lam:2017ofc,Banerjee:2017jeg,Stieberger:2018edy}. In this section we calculate the modified Mellin transform of some low-point tree level amplitudes for gravitons and gluons.   

\subsection{Four particle graviton tree amplitude in Einstein gravity}\label{4gravMHV}
It is known that the tree-level Mellin amplitude $\tilde S$ for gravitons is divergent in Einstein gravity \cite{Stieberger:2018edy}. An interesting resolution suggested by \cite{Stieberger:2018edy} is that instead of Einstein gravity one should consider graviton scattering in string theory and this indeed makes the amplitude $\tilde S$ finite. However, it seems that there is no limit in which one can recover the classical Einstein gravity from the $\tilde S$ amplitude computed in string theory. In this paper we show that the modified Mellin amplitude $\mathcal A$ is in fact finite for tree level graviton scattering amplitude in Einstein gravity. It turns out that separating the asymptotic particles in retarded time naturally regularizes the $\mathcal A$ amplitudes in Einstein gravity. Let us now describe the results. 

With gravitons $1,2$ of negative helicities and $3,4$ of positive helicities the four graviton tree amplitude in spinor-helicity variables is 
\be\label{gmhv}
S(1^{--}2^{--}3^{++}4^{++})(\om_i,z_i,\bar{z}_i)=\frac{\kappa^{2}}{4} \frac{\langle 12 \rangle^4 [34]^4}{s t u}\delta^4 (\sum_i \varepsilon_i \om_i q_i)~~
\ee
where $\kappa=\sqrt{8 \pi G_{N}}$ and $G_{N}$ is the $4$-dimensional Newton's constant. 

We define the kinematic variables $s,t,u$ as
\be
s=\langle 12 \rangle [12]~~;~~
t=\langle 13 \rangle [13]~~;~~
u=\langle 14 \rangle [14]
\ee
Using the definition of spinor helicity variables \cite{Pasterski:2017ylz} and the parametrisation of null momentum as in \eqref{para}, we have,
\be
	\langle ij\rangle=- \varepsilon_i \varepsilon_j 2\sqrt{\om_i \om_j}z_{ij}~~;~~
\lbrack ij \rbrack=2\sqrt{\om_i \om_j}\bar{z}_{ij} \quad ; \quad z_{ij}=z_i -z_j
\ee
So in terms of $\omega,z,\bar{z}$ the $S$-matrix \eqref{gmhv} becomes,
\be\label{gmhv1}
S(1^{--}2^{--}3^{++}4^{++})(\om_i,z_i,\bar{z}_i)=\kappa^{2}\frac{\om_2 \om_3\om_4 z_{12}^4 \bar{z}^4_{34}}{\om_1 z_{12} \bar{z}_{12} z_{13}\bar{z}_{13}z_{14}\bar{z}_{14} }\delta^4 (\sum_i \varepsilon_i \om_i q_i)
\ee
Now we want to find out the modified amplitude $\mathcal A$ corresponding to \eqref{gmhv1}. Before doing this it is convenient to change $\omega_i$'s to the simplex variables \cite{Pasterski:2017ylz} :  $\sigma_i=s^{-1}\om_i $ , $\sum_i \om_i =s$. In terms of simplex variables $\mathcal A$ can be written as,
 

\bea
\mathcal{A}(1^{--}2^{--}3^{++}4^{++})(u_i,z_i,\bar{z}_i,\l_i)\nn =
\prod_{i=1} ^4 \int_{0} ^{\infty}d\om_i \om^{i \l_i}e^{-i\varepsilon_i \om_i u_i}\lbrack ...\rbrack \\ =\int_{0}^{\infty}ds~ s^{4-1+i\sum_i \l_i} \prod_{i=1} ^4 \int_{0} ^{1}d\sigma_i \sigma_i^{i\l_i} e^{-i\varepsilon_i s\sigma_i u_i}\delta(\sum_i \sigma_i -1)\lbrack ...\rbrack\nn\\
\label{eq2.6}
\eea
where,
\bea
[...] = S(1^{--}2^{--}3^{++}4^{++})(s,\s_i,z_i,\bar{z}_i)=s^{-2}\kappa^{2}\frac{\s_2 \s_3\s_4 z_{12}^3 \bar{z}^4_{34}}{\s_1 z_{13}\bar{z}_{13}z_{14}\bar{z}_{14}\bar{z}_{12}}\delta^4 (\sum_i \varepsilon_i \s_i q_i)
\eea
We can rewrite the delta functions as \cite{Pasterski:2017ylz},
\bea
\delta^4 (\sum_i \varepsilon_i\sigma_i q_i)\delta(\sum_i\sigma_i -1)=C(z_i,\bar{z}_i)\prod_{i=1}^{4}\delta(\sigma_i-\sigma_i^\ast)\nn\\
C(z_i,\bar{z}_i)=\frac{1}{4}\delta\(|z_{12}z_{34}\bar{z}_{13}\bar{z}_{24}-\bar{z}_{12}\bar{z}_{34}z_{13}z_{24}| \)=\frac{\delta(|z-\bar{z}|)}{4 z_{13}\bar{z}_{13}z_{24}\bar{z}_{24}}\nn\\
\sigma_1^\ast =-\frac{\varepsilon_1 \varepsilon_4}{D}\frac{z_{24}\bar{z}_{34}}{z_{12}\bar{z}_{13}}~~,~~
\sigma_2^\ast =\frac{\varepsilon_2 \varepsilon_4}{D}\frac{z_{34}\bar{z}_{14}}{z_{23}\bar{z}_{12}}\nn\\
\sigma_3^\ast =-\frac{\varepsilon_3 \varepsilon_4}{D}\frac{z_{24}\bar{z}_{14}}{z_{23}\bar{z}_{13}}~~,~~
\sigma_4^\ast =\frac{1}{D}\nn\\
D=(1-\varepsilon_1\varepsilon_4)\frac{z_{24}\bar{z}_{34}}{z_{12}\bar{z}_{13}}+(\varepsilon_2\varepsilon_4-1)\frac{z_{34}\bar{z}_{14}}{z_{23}\bar{z}_{12}}+(1-\varepsilon_3\varepsilon_4)\frac{z_{24}\bar{z}_{14}}{z_{23}\bar{z}_{13}}
\label{eq2.5}
\eea

In terms of these the modified Mellin amplitude $\mathcal{A}$ becomes,

\begin{gather}\label{di}
\mathcal{A}(1^{--}2^{--}3^{++}4^{++})(u_i,z_i,\bar{z}_i,\l_i)\nn\\
=\int_{0}^{\infty}ds~ s^{4-1+i\sum_i \l_i} \prod_{i=1} ^4 \int_{0} ^{1}d\sigma_i \sigma_i^{i\l_i} e^{-i\varepsilon_i s\sigma_i u_i}\delta(\sum_i \sigma_i -1)S(1^{--}2^{--}3^{++}4^{++})(\s_i,z_i,\bar{z}_i)\nn\\
=\frac{\kappa^{2}}{4}\(\int_{0}^{\infty}ds ~s^{1+i\sum_i \l_i}e^{-i\sum_i\varepsilon_i s\sigma_i^\ast u_i}\)\(\prod_i \sigma_i^{\ast i \l_i}\) \frac{\s_2^\ast \s_3^\ast \s_4^\ast z_{12}^3 \bar{z}^4_{34}}{\s_1^\ast z_{13}\bar{z}_{13}z_{14}\bar{z}_{14}\bar{z}_{12}}\frac{\delta(|z-\bar{z}|)}{z_{13}\bar{z}_{13}z_{24}\bar{z}_{24}} \prod_{i=1}^{4}\textbf{1} _{[0,1]}(\s_i^\ast)\label{eq12} \nn \\
= \lim_{\delta\rightarrow 0+} \frac{\kappa^{2}}{4}\(\int_{0}^{\infty}ds ~s^{1+i\sum_i \l_i}e^{-i\sum_i\varepsilon_i s\sigma_i^\ast u_i}e^{-\delta s}\)\(\prod_i \sigma_i^{\ast i \l_i}\) \frac{\s_2^\ast \s_3^\ast \s_4^\ast z_{12}^3 \bar{z}^4_{34}}{\s_1^\ast z_{13}\bar{z}_{13}z_{14}\bar{z}_{14}\bar{z}_{12}}\frac{\delta(|z-\bar{z}|)}{z_{13}\bar{z}_{13}z_{24}\bar{z}_{24}} \prod_{i=1}^{4}\textbf{1} _{[0,1]}(\s_i^\ast)\label{eq12}
\end{gather}
Here we have regularized the integral as described in section-\eqref{reg}. The $\s_i$ integrals are done by simply using the delta functions. Here the term $\prod_{i=1}^{4}\textbf{1} _{[0,1]}(\s_i^\ast)$ ensures that $\s_i ^\ast$ are in range $[0,1]$.

\be
\textbf{1} _{[0,1]}(\s_i^\ast)=\begin{cases}
	1, & \s_i ^\ast ~\epsilon ~[0,1]\\
	0, & otherwise
\end{cases}
\ee

Now using
\be
\lim_{\delta\rightarrow 0+}\int_{0}^{\infty}ds ~s^{1+i\sum_i \l_i}e^{-i\sum_i\varepsilon_i s\sigma_i^\ast u_i} e^{-\delta s}= \lim_{\delta\rightarrow 0+}\frac{\Gamma (2+i\sum \l_i)}{(i \sum \varepsilon_i \sigma_i ^\ast u_i + \delta)^{2+i\sum \l_i}}
\ee
the final result for the modified Mellin amplitude becomes,


\be\label{fmm}
\begin{gathered}
\mathcal{A}(1^{--}2^{--}3^{++}4^{++})(u_i,z_i,\bar{z}_i,\l_i)\\
=\frac{\kappa^{2}}{4}\bigg[\lim_{\delta\rightarrow 0+}\frac{\Gamma (2+i\sum \l_i)}{(i \sum \varepsilon_i \sigma_i ^\ast u_i + \delta)^{2+i\sum \l_i}}\bigg]\(\prod_i \sigma_i^{\ast i \l_i}\) \frac{\s_2^\ast \s_3^\ast \s_4^\ast z_{12}^3 \bar{z}^4_{34}}{\s_1^\ast z_{13}\bar{z}_{13}z_{14}\bar{z}_{14}\bar{z}_{12}} \frac{\delta(|z-\bar{z}|)}{ z_{13}\bar{z}_{13}z_{24}\bar{z}_{24}}\prod_{i=1}^{4}\textbf{1} _{[0,1]}(\s_i^\ast)
\end{gathered}
\ee

One can check that the amplitude \eqref{fmm} is translationally invariant. Under global space-time translation, $u_i \rightarrow u_i + A + Bz_i + \bar B \bar z_i + C z_i\bar z_i$, and the change in the term $\sum_{i=1}^{4}\varepsilon_i \s_i^\ast u_i$ is always proportional to $(z-\bar z)$. Hence the amplitude is invariant due to the $\delta(|z-\bar{z}|)$ constraint. 

\subsection{Four particle gluon tree amplitude in Yang-Mills} 
Four particle gluon amplitude for particles $1,2$ with negative helicity and $3,4$ with positive helicity is given as,
\be
S(1^- 2^- 3^+ 4^+)(\omega_i ,z_i ,\bar{z}_i)=\frac{\langle 12 \rangle^3}{\langle 23\rangle \langle 34\rangle \langle 41\rangle}\delta^4(\sum_{i=1}^{4}\varepsilon_i \omega_i q_i)
\ee	
Writing this amplitude in terms of $\omega,z,\bar{z}$ we get,
\be
S(1^- 2^- 3^+ 4^+)(\omega_i ,z_i ,\bar{z}_i)=\frac{\omega_1 \omega_2}{\om_3\om_4}\frac{z_{12}^3}{z_{23}z_{34}z_{41}}\delta^4(\sum_{i=1}^{4}\varepsilon_i \omega_i q_i)
\ee
Rewriting this amplitude in terms of simplex variables, 
\be
S(1^{-}2^{-}3^{+}4^{+})(\s_i,z_i,\bar{z}_i)=s^{-4}\frac{\s_1 \s_2}{\s_3\s_4}\frac{z_{12}^3}{z_{23}z_{34}z_{41}}\delta^4(\sum_{i=1}^{4}\varepsilon_i \s_i q_i)
\ee
Using \eqref{eq2.5} and \eqref{eq2.6} we find the modified Mellin amplitude for four gluons to be,	
\begin{gather}
\mathcal{A}(1^- 2^- 3^+ 4^+)(u_i ,z_i,\bar{z}_i,\l_i)\nn\\=\int_{0}^{\infty}ds~ s^{4-1+i\sum_i \l_i} \prod_{i=1} ^4 \int_{0} ^{1}d\sigma_i \sigma_i^{i \l_i} e^{-i\varepsilon_i s\sigma_i u_i}\delta(\sum_i \sigma_i -1)S(1^{-}2^{-}3^{+}4^{+})(\s_i,z_i,\bar{z}_i)\nn\\
=\frac{1}{4}\left(\int_{0}^{\infty}ds s^{-1+i\sum_{i=1}^4\l_i}e^{-i\sum_{i=1}^{4}\varepsilon_i s\sigma_i^\ast u_i}\right)\(\prod_i \sigma_i^{\ast i \l_i}\)\frac{\s_1^\ast \s_2^\ast}{\s_3^\ast \s_4^\ast}\frac{z_{12}^3}{z_{23}z_{34}z_{41}}\frac{\delta(|z-\bar{z}|)}{ z_{13}\bar{z}_{13}z_{24}\bar{z}_{24}}\prod_{i=1}^{4}\textbf{1} _{[0,1]}(\s_i^\ast)
\end{gather}
The $u_i$ dependence of the amplitude is given by,
\be
\int_{0}^{\infty}ds s^{-1+i\sum_{i=1}^4 \l_i}e^{-i\sum_{i=1}^{4}\varepsilon_i s\sigma_i^\ast u_i}=\lim_{\delta\rightarrow 0+}\frac{\Gamma(i\sum_i \l_i)}{\left(i\sum_{i=1}^{4}\varepsilon_i\sigma_i^\ast u_i + \delta \right)^{i\sum_i \l_i}}
\ee
The final result is given by,

\begin{eqnarray}\label{gluon-4pt}
&&\mathcal{A}(1^- 2^- 3^+ 4^+)(u_i ,z_i,\bar{z}_i,\l_i)\nn\\ &= &\frac{1}{4}\bigg[ \lim_{\delta\rightarrow 0+}\frac{\Gamma(i\sum_i \l_i)}{\left(i\sum\varepsilon_i \sigma_i^\ast u_i +\delta  \right)^{i\sum_i \l_i}} \bigg]\(\prod_i \sigma_i^{\ast i \l_i}\)\frac{\s_1^\ast \s_2^\ast}{\s_3^\ast \s_4^\ast}\frac{z_{12}^3}{z_{23}z_{34}z_{41}}\frac{\delta(|z-\bar{z}|)}{ z_{13}\bar{z}_{13}z_{24}\bar{z}_{24}}\prod_{i=1}^{4}\textbf{1} _{[0,1]}(\s_i^\ast) \nn\\
\end{eqnarray}
Just like in the graviton case this amplitude is also translationally invariant due to the Dirac delta function constraint on the cross ratios.  


\section{Conformal Soft Factorization}\label{csf}

Let us now show that the modified amplitudes admit \emph{conformal} soft factorization \cite{Fan:2019emx,Nandan:2019jas,Pate:2019mfs,Donnay:2018neh,Adamo:2019ipt,Puhm:2019zbl,Guevara:2019ypd}.

\subsection{$n$-point graviton amplitudes }

Consider a $n$-point graviton scattering amplitude $S_{n} \left(\omega_{i},z_{i},\bar{z}_{i},\sigma_i\right)$. Now if we take the limit where the momentum of the $n$-th particle goes to zero, then according to Weinberg's soft graviton theorem \cite{Weinberg:1965nx} the $n$-point amplitude factorizes into a $(n-1)$ point amplitude as follows

\begin{equation}
\label{softgrav}
\begin{split}
S_{n} \left(\omega_{i},z_{i},\bar{z}_{i},\sigma_i\right)= \mathcal{F}^{(0)} S_{n-1} \left(\omega_{i},z_{i},\bar{z}_{i},\sigma_i\right) + \ldots
\end{split}
\end{equation}

where $\mathcal{F}^{(0)} $ is the universal leading Weinberg soft factor. The ellipses in equation \ref{softgrav} denote subleading contributions in the soft momentum. However, in this section we will only concern ourselves with the leading soft factor. 

Now if the soft graviton has positive helicity,  the soft factor can be expressed in terms of spinor helicity variables as  \cite{Cachazo:2014fwa} 

\begin{equation}
\label{softfactor}
\begin{split}
\mathcal{F}^{(0)}_{+} = -\sum_{i=1}^{n-1} \frac{[n i]}{\langle n i \rangle} \frac{\langle x i\rangle \langle y i \rangle}{ \langle x n \rangle \langle y n \rangle}
\end{split}
\end{equation}

Here we have introduced a subscript $(+)$ in the soft factor to explicitly denote the case that the soft graviton here has positive helicity. Now using the parametrisation of null momenta in equation \ref{para} this becomes

\begin{equation}
\label{softfactor1}
\begin{split}
\mathcal{F}^{(0)}_{+} = -\sum_{i=1}^{n-1} \frac{\varepsilon_{i}\omega_{i}}{\varepsilon_{n} \omega_{n}} \frac{\bar{z}_{ni}z_{x i}z_{yi}}{z_{ni}z_{xn}z_{yn}}
\end{split}
\end{equation}

Similarly in the case where the soft graviton has negative helicity, we have

\begin{equation}
\label{softfacantihol}
\begin{split}
\mathcal{F}^{(0)}_{-} = -\sum_{i=1}^{n-1} \frac{\langle n i\rangle}{[n i ]} \frac{[ x i ][ y i ]}{ [x n] [ y n]}= - \sum_{i=1}^{n-1}   \frac{\varepsilon_{i}\omega_{i}}{\varepsilon_{n} \omega_{n}}  \frac{z_{ni}}{\bar{z}_{ni}}\frac{\bar{z}_{x i}\bar{z}_{yi}}{\bar{z}_{xn}\bar{z}_{yn}}
\end{split}
\end{equation}

Let us now consider the conformal soft limit of the modified Mellin transform of the $n$-point graviton amplitude. Following \cite{Fan:2019emx,Nandan:2019jas,Pate:2019mfs}, we define this as,

\begin{equation}
\label{confsoftgrav0}
\begin{split}
& \lim_{\lambda_{n} \to 0} i \lambda_{n} \hspace{0.04cm}\mathcal{A}_{n} (u_{i},z_{i},\bar{z}_{i},\lambda_{i},\sigma_i)
\end{split}
\end{equation} 

Recently in \cite{Puhm:2019zbl} the conformal soft limit of graviton amplitudes has been studied. It is important to note that the main difference between our analysis and the analysis in \cite{Puhm:2019zbl} is that we are using the prescription \eqref{modmellin} of \cite{Banerjee:2018gce,Banerjee:2018fgd} for defining the Mellin transform of scattering amplitudes which is a modification of the definition proposed in \cite{Pasterski:2017ylz} and subsequently used in \cite{Puhm:2019zbl}. Now, using our definition for the Mellin transform in equation \eqref{modmellin} we get

\begin{equation}
\label{conformalsoftgrav0}
\begin{split}
& \lim_{\lambda_{n} \to 0} i \lambda_{n} \hspace{0.04cm}\mathcal{A}_{n} (u_{i},z_{i},\bar{z}_{i},\lambda_{i},\sigma_i)\\
&  = \int \prod_{k=1}^{n-1} d\omega_{k} \hspace{0.05cm} \omega_{k}^{i \lambda_{k}} \hspace{0.1cm} e^{-i \sum\limits_{k=1}^{n-1}\varepsilon_{k}\omega_{k}u_{k}} \int_{0}^{\infty} d\omega_{n} \left( \lim_{\lambda_{n} \to 0} i \lambda_{n}\hspace{0.04cm}\omega_{n}^{i\lambda_{n}-1}\right)\hspace{0.04cm} \omega_{n} \hspace{0.04cm} e^{- i \varepsilon_{n}\omega_{n}u_{n}}S_{n} \left(\omega_{i},z_{i},\bar{z}_{i},\sigma_i\right)\\
& = \int \prod_{k=1}^{n-1} d\omega_{k} \hspace{0.05cm} \omega_{k}^{i \lambda_{k}} \hspace{0.1cm} e^{-i \sum\limits_{k=1}^{n-1}\varepsilon_{k}\omega_{k}u_{k}} \int_{0}^{\infty} d\omega_{n}\hspace{0.04cm} \delta(\omega_{n})\hspace{0.04cm} \omega_{n} \hspace{0.04cm} e^{- i \varepsilon_{n}\omega_{n}u_{n}} S_{n} \left(\omega_{i},z_{i},\bar{z}_{i},\sigma_i\right)\\
\end{split}
\end{equation}

where in the last line above we have used the identity 
\begin{equation}
\label{deltafuncid}
\begin{split}
& \delta (x) = \frac{1}{2}  \lim_{\epsilon \to 0 } \epsilon \hspace{0.04cm} |x|^{\epsilon-1}
\end{split}
\end{equation}

Using this limit representation of the delta function inside the $\omega_{n}$ integral is well defined only if remaining integrand does not diverge for large $\omega_{n}$. In our case this is a valid operation since as we argued in section \eqref{reg}, the factor $e^{-i\varepsilon_n \omega_{n}u_{n}}$ guarantees convergence of the integral for large $\omega_{n}$ upon analytically continuing $u_{n}$ to $u_{n}-i \varepsilon_{n} \delta $. Here $\delta\rightarrow 0^{+}$ and $\varepsilon_{n}=\pm 1$ if the $n$-th particle is outgoing (incoming). \\

Now using  \eqref{softgrav} in  equation \eqref{conformalsoftgrav0} we get
\begin{equation}
\label{conformalsoftgrav}
\begin{split}
\lim_{\lambda_{n} \to 0} i \lambda_{n} \hspace{0.04cm}\mathcal{A}_{n} (u_{i},z_{i},\bar{z}_{i},\lambda_{i},\sigma_i)& = - \int \prod_{k=1}^{n-1} d\omega_{k} \hspace{0.05cm} \omega_{k}^{i \lambda_{k}} \hspace{0.1cm} e^{-i \sum\limits_{k=1}^{n-1}\varepsilon_{k}\omega_{k}u_{k}}  \hspace{0.06cm} \sum_{i=1}^{n-1}\left( \frac{\varepsilon_{i}\omega_{i} }{\varepsilon_{n}} \frac{\bar{z}_{ni}z_{x i}z_{yi}}{z_{ni}z_{xn}z_{yn}} \right)S_{n-1} \left(\omega_{i},z_{i},\bar{z}_{i},\sigma_i\right) \\
& = -  i \varepsilon_{n} \sum_{j=1}^{n-1} \frac{\bar{z}_{ni}z_{x i}z_{yi}}{z_{ni}z_{xn}z_{yn}} \frac{\partial}{\partial u_{j}} \int \prod_{k=1}^{n-1} d\omega_{k} \hspace{0.05cm} \omega_{k}^{i \lambda_{k}} \hspace{0.1cm} e^{-i \sum\limits_{k=1}^{n-1}\varepsilon_{k}\omega_{k}u_{k}}  \hspace{0.06cm} S_{n-1} \left(\omega_{i},z_{i},\bar{z}_{i},\sigma_i\right)\\
&= - \left(  i  \varepsilon_{n} \sum_{i=1}^{n-1}   \frac{\bar{z}_{ni}}{z_{ni}}\frac{z_{x i}z_{yi}}{z_{xn}z_{yn}}\frac{\partial}{\partial u_{i}} \right) \mathcal{A}_{n-1} (u_{i},z_{i},\bar{z}_{i},\lambda_{i},\sigma_i)
\end{split}
\end{equation}

where in the second line above we have used,
\begin{equation}
\label{conformalsoftgrav1}
\begin{split}
\omega_{j} e^{-i \varepsilon_{j} \omega_{j}u_{j}} = i \varepsilon_{j} \frac{\partial}{\partial u_{j}} \left(  e^{-i \varepsilon_{j} \omega_{j}u_{j}} \right)
\end{split}
\end{equation}

Thus in the conformal soft limit the Mellin transformed graviton amplitude takes the form

\begin{equation}
\label{conformalsoftgrav2}
\begin{split}
\lim_{\lambda_{n} \to 0} i \lambda_{n}\hspace{0.04cm} \mathcal{A}_{n} (u_{i},z_{i},\bar{z}_{i},\lambda_{i},\sigma_i)&= \tilde{\mathcal{F}}^{(0)}_{+} \mathcal{A}_{n-1} (u_{i},z_{i},\bar{z}_{i},\lambda_{i},\sigma_i)
\end{split}
\end{equation}

where we have defined the conformal soft factor $ \tilde{\mathcal{F}}^{(0)}_{+} $ to be
\begin{equation}
\label{confsoftfachol}
\begin{split}
\tilde{\mathcal{F}}^{(0)}_{+} = -   i  \varepsilon_{n} \sum_{i=1}^{n-1}   \frac{\bar{z}_{ni}}{z_{ni}}\frac{z_{x i}z_{yi}}{z_{xn}z_{yn}}\frac{\partial}{\partial u_{i}} 
\end{split}
\end{equation}

Similarly for a negative helicity soft graviton, we obtain

\begin{equation}
\label{conformalsoftgrav3}
\begin{split}
\lim_{\lambda_{n} \to 0} i \lambda_{n}\hspace{0.04cm} \mathcal{A}_{n} (u_{i},z_{i},\bar{z}_{i},\lambda_{i},\sigma_i)&= \tilde{\mathcal{F}}^{(0)}_{-} \mathcal{A}_{n-1} (u_{i},z_{i},\bar{z}_{i},\lambda_{i},\sigma_i)
\end{split}
\end{equation}

where
\begin{equation}
\label{confsoftfacantihol}
\begin{split}
\tilde{\mathcal{F}}^{(0)}_{-} = -   i  \varepsilon_{n} \sum_{i=1}^{n-1}   \frac{z_{ni}}{\bar{z}_{ni}}\frac{\bar{z}_{x i}\bar{z}_{yi}}{\bar{z}_{xn}\bar{z}_{yn}}\frac{\partial}{\partial u_{i}} 
\end{split}
\end{equation}

Now, supertranslation acts naturally on the modified amplitude $\mathcal A(u_i,z_i,\bar z_i,\sigma_i)$ by shifting $u_i$ to $u_i + f(z_i,\bar z_i)$. The appearance of the derivative $\partial/\partial u$ is related to the fact that the leading soft graviton theorem is essentially the Ward identity for supertranslation \cite{Strominger:2013jfa}. In \cite{Puhm:2019zbl} the analog of $\partial/\partial u$, acting on the lower point amplitude, is the shift of each $\lambda$ to $\lambda - i$ in the lower point amplitude appearing on the R.H.S of the conformal soft theorem for gravitons. 

In the next subsection we will explicitly check this general result in the case of $4$-point MHV graviton amplitudes. For this let us first compute the modified Mellin transform of $3$-point graviton amplitudes which will now appear on the R.H.S. of the soft theorem. 

\subsection{$3$-point MHV amplitude}\label{MHVgrav}

Since $3$-point graviton amplitudes vanish in ordinary Minkowski signature, in this section we will work with the mixed signature spacetime metric $(-,+,-,+)$. In this case the parametrisation of null momenta becomes

\begin{equation}
\label{nullmom2}
\begin{split}
p_i = \omega_i (1+z_i\bar z_i, z_i + \bar z_i , z_i - \bar z_i, 1- z_i \bar z_i)
\end{split}
\end{equation}

where $z_{i}, \bar{z}_{i}$ are now independent real numbers. The $3$-point graviton MHV amplitude is then given by

\begin{equation}
\begin{split}
S(1^{--},2^{--},3^{++})(\omega_{i},z_{i},\bar{z}_{i})&=\frac{\kappa}{2} \frac{\langle 12\rangle^{6}}{\langle 23\rangle^{2} \langle 31\rangle^{2}}\delta^{(4)}\left(\sum_{i=1}^{3}\epsilon_{i}p_{i}^{\mu}\right)\\
&= 2 \kappa \hspace{0.06cm} \frac{\omega_{1}^{2}\omega_{2}^{2}}{\omega_{3}^{2}}\frac{z_{12}^{6}}{z_{23}^{2}z_{31}^{2}}\delta^{(4)}\left(\sum_{i=1}^{3}\epsilon_{i}\omega_{i}q_{i}^{\mu}\right)
\end{split}
\end{equation}

As in the case of $4$-point amplitudes, it is convenient to first define the simplex variables, $\sigma_{i}= s^{-1} \omega_{i}, s=\sum_{i}\omega_{i}$ before computing the Mellin transform of the amplitude. Then we can write the momentum conservation imposing delta function as \cite{Pasterski:2017ylz}

\begin{equation}
\label{momentadel}
\begin{split}
\delta^{(4)}\left(\sum_{i=1}^{3}\epsilon_{i}\omega_{i}q_{i}^{\mu}\right)= s^{-4}\hspace{0.05cm}\delta^{(4)}\left(\sum_{i=1}^{3}\epsilon_{i}\sigma_{i}q_{i}^{\mu}\right)\delta\left(\sum_{i=1}^{3}\sigma_{i}-1\right)= \frac{\delta(z_{12})\delta(z_{13})}{4 \sigma_{1}\sigma_{2}\sigma_{3}D_{3}^{2}} \prod_{i=1}^{3}\delta(\sigma_{i}-\sigma^{*}_{i})
\end{split}
\end{equation}

where
\begin{equation}
\label{sigma3ptmhv}
\begin{split}
& \sigma_{1}^{*} = \frac{z_{23}}{D_{3}} , \quad \sigma_{2}^{*} =-\varepsilon_{1}\varepsilon_{2} \frac{z_{13}}{D_{3}}, \quad \sigma_{3}^{*} =\varepsilon_{1}\varepsilon_{3} \frac{z_{12}}{D_{3}} \\
& D_{3}=(1-\varepsilon_{1}\varepsilon_{2})z_{13} + (\varepsilon_{1}\varepsilon_{3}-1) z_{12}
\end{split}
\end{equation}

Now the modified Mellin transform of the $3$-point MHV amplitude is 

\begin{equation}
\begin{split}
& \mathcal{A}(1^{--},2^{--},3^{++})(u_{i},z_{i},\bar{z}_{i},\lambda_{i})\\
&=\prod_{i=1}^{3}\int d\omega_{i} \hspace{0.05cm}\omega_{i}^{i \lambda_{i}}\hspace{0.05cm}e^{-i\epsilon_{i}\omega_{i}u_{i}}S(1^{--},2^{--},3^{++})(\omega_{i},z_{i},\bar{z}_{i})\\
&=2 \hspace{0.04cm}\kappa \int \prod_{i=1}^{3} d\sigma_{i} \hspace{0.05cm}\sigma_{i}^{i \lambda_{i}}\hspace{0.05cm}\int ds \hspace{0.05cm} s^{i\sum\limits_{i}\lambda_{i}}e^{-is\sum\limits_{i}\epsilon_{i}\sigma_{i}u_{i}}\hspace{0.05cm}\frac{\sigma_{1}^{2}\sigma_{2}^{2}}{\sigma_{3}^{2}}\frac{z_{12}^{6}}{z_{23}^{2}z_{31}^{2}}\hspace{0.05cm}\delta^{(4)}\left(\sum_{i=1}^{3}\epsilon_{i}\sigma_{i}q_{i}^{\mu}\right)\delta\left(\sum_{i=1}^{3}\sigma_{i}-1\right)
\end{split}
\end{equation}

The integral over the energy scale $s$ yields, 
\begin{equation}
\int ds \hspace{0.05cm} s^{i\sum_{i}\lambda_{i}}e^{-is\sum_{i}\epsilon_{i}\sigma_{i}u_{i}}= \left[\lim_{\delta\rightarrow 0+}\frac{\Gamma (1+i\Lambda)}{\left(i \sum \varepsilon_i \sigma_i u_i + \delta\right)^{1+i\Lambda}}\right] 
\end{equation}

where $\Lambda = \sum\limits_{i=1}^{3}\lambda_{i}$. Then performing the $\sigma_{i}$ integrals using the representation of the delta functions in equation \eqref{momentadel} we get

\begin{equation}
\begin{split}
\label{mellin3gravmhv}
& \mathcal{A}(1^{--},2^{--},3^{++})(u_{i},z_{i},\bar{z}_{i},\lambda_{i})\\
&=  \hspace{0.04cm}\kappa \left[\lim_{\delta\rightarrow 0+}\frac{\Gamma (1+i\Lambda)}{\left(i \sum \varepsilon_i \sigma^{*}_{i} u_i + \delta\right)^{1+i\Lambda}}\right]  \prod_{i=1}^{3}(\sigma_{i}^{*})^{i \lambda_{i}}\left(\frac{\sigma_{1}^{*}\sigma_{2}^{*}}{\sigma_{3}^{*}}\frac{z_{12}^{3}}{z_{23}z_{31}}\right)^{2}\frac{\delta(\bar{z}_{12})\delta(\bar{z}_{13})}{2\sigma_{1}^{*}\sigma_{2}^{*}\sigma_{3}^{*}D_{3}^{2}} \prod_{i=1}^{3} \mathbf{1}_{[0,1]}\left(\sigma^{*}_{i}\right)\\
\end{split}
\end{equation}

where the indicator function $ \prod\limits_{i=1}^{3} \mathbf{1}_{[0,1]}\left(\sigma^{*}_{i}\right)$ imposes the constraint that $\sigma_{i}^{*}\in [0,1]$.

\subsection{$3$-point $\overline{\mathrm{MHV}}$ amplitude}\label{antimhvgrav}

Here we will compute the modified Mellin transform of the $3$-point graviton $\overline{\mathrm{MHV}}$ amplitude. In momentum space this is given by
\begin{equation}
\begin{split}
S(1^{++},2^{++},3^{--})(\omega_{i},z_{i},\bar{z}_{i})&=\frac{\kappa}{2} \frac{[12]^{6}}{[23]^{2}[31]^{2}}\delta^{(4)}\left(\sum_{i=1}^{3}\epsilon_{i}p_{i}^{\mu}\right)\\
&= 2 \hspace{0.04cm} \kappa \hspace{0.06cm}\frac{\omega_{1}^{2}\omega_{2}^{2}}{\omega_{3}^{2}}\frac{\bar{z}_{12}^{6}}{\bar{z}_{23}^{2}\bar{z}_{31}^{2}}\delta^{(4)}\left(\sum_{i=1}^{3}\epsilon_{i}\omega_{i}q_{i}^{\mu}\right)
\end{split}
\end{equation}

Now the modified Mellin transform can be obtained following essentially identical steps as for the MHV amplitude in the previous section.  The final result is simply given by equation \eqref{mellin3gravmhv} with the subsitiution $z_{ij} \rightarrow \bar{z}_{ij}$ as follows

\begin{equation}
\begin{split}
\label{mellin3gravbar}
& \mathcal{A}(1^{++},2^{++},3^{--})(u_{i},z_{i},\bar{z}_{i},\lambda_{i})\\
&=  \hspace{0.04cm} \kappa \left[\lim_{\delta\rightarrow 0+}\frac{\Gamma (1+i\Lambda)}{\left(i \sum \varepsilon_i \sigma^{*}_{i} u_i + \delta\right)^{1+i\Lambda}}\right]  \prod_{i=1}^{3}(\sigma_{i}^{*})^{i \lambda_{i}}\left(\frac{\sigma_{1}^{*}\sigma_{2}^{*}}{\sigma_{3}^{*}}\frac{\bar{z}_{12}^{3}}{\bar{z}_{23}{z}_{31}}\right)^{2}\frac{\delta(z_{12})\delta(z_{13})}{2\sigma_{1}^{*}\sigma_{2}^{*}\sigma_{3}^{*}D_{3}^{2}} \prod_{i=1}^{3} \mathbf{1}_{[0,1]}\left(\sigma^{*}_{i}\right)\\
\end{split}
\end{equation}

where now we have
\begin{equation}
\label{sigma3ptantimhv}
\begin{split}
& \sigma_{1}^{*} = \frac{\bar{z}_{23}}{D_{3}} , \quad \sigma_{2}^{*} =-\varepsilon_{1}\varepsilon_{2} \frac{\bar{z}_{13}}{D_{3}}, \quad \sigma_{3}^{*} =\varepsilon_{1}\varepsilon_{3} \frac{\bar{z}_{12}}{D_{3}} \\
& D_{3}=(1-\varepsilon_{1}\varepsilon_{2})\bar{z}_{13} + (\varepsilon_{1}\varepsilon_{3}-1) \bar{z}_{12}
\end{split}
\end{equation}

\subsection{Conformal soft limit of MHV graviton amplitudes}

We can now verify the conformal soft factorization result obtained in equations \eqref{conformalsoftgrav2} and \eqref{conformalsoftgrav3} for $4$-point tree-level MHV graviton amplitudes.  Our analysis in this section will be similar to the one carried out in \cite{Puhm:2019zbl}. Here again we will take the signature of the spacetime metric to be $(-,+,-,+)$. 

The modified Mellin transform of the $4$-point MHV amplitude was computed in section \eqref{4gravMHV} and is given by 

\begin{equation}
\label{mellinmhvgrav}
\begin{split}
& \mathcal{A}(1^{--},2^{--},3^{++},4^{++})(u_i,z_i,\bar{z}_i,\l_i)\\
&=\frac{\kappa^{2}}{4}\bigg[\lim_{\delta\rightarrow 0+}\frac{\Gamma (2+i\Lambda)}{(i \sum \varepsilon_i \sigma_i ^\ast u_i + \delta)^{2+i\Lambda}}\bigg]\(\prod_{i=1}^{4} \sigma_i^{\ast i \l_i}\) \frac{\s_2^\ast \s_3^\ast \s_4^\ast }{\s_1^\ast }  f(z_{i},\bar{z}_{i}) \prod_{i=1}^{4}\textbf{1} _{[0,1]}(\s_i^\ast)
\end{split}
\end{equation}

where we have defined
\begin{equation}
\label{fzzbar}
\begin{split}
f(z_{i},\bar{z}_{i}) = \frac{z_{12}^{3}\bar{z}_{34}^{4}}{\bar{z}_{12} z_{13}\bar{z}_{13}z_{14}\bar{z}_{14}} \delta\left(z_{12}z_{34}\bar{z}_{13}\bar{z}_{24}- z_{13}z_{24}\bar{z}_{12}\bar{z}_{34}\right), \quad\quad \Lambda= \sum_{i=1}^{4}\lambda_{i}
\end{split}
\end{equation}

with $z_{i},\bar{z}_{i}$ being independent real quantities. Now let us consider the conformal soft limit of the Mellin amplitude in equation \eqref{mellinmhvgrav} by taking the $1$-st negative helicity particle to be soft. Thus we have

\begin{equation}
\label{confsoftmhv}
\begin{split}
& \lim_{\lambda_{1}\to 0} i \lambda_{1} \hspace{0.05cm} \mathcal{A}(1^{--},2^{--},3^{++},4^{++})(u_i,z_i,\bar{z}_i,\l_i)\\
&= \frac{\kappa^{2}}{4} \left(\lim_{\lambda_{1} \to 0} i \lambda_{1} \left[\lim_{\delta\rightarrow 0+}\frac{\Gamma (2+i\Lambda)}{\left(i \sum \varepsilon_i \sigma_i ^\ast u_i + \delta\right)^{2+i\Lambda}}\right] \sigma_{1}^{\ast i \lambda_{1}-1} \right) \left(\prod_{i=2}^{4} \sigma_{i}^{\ast 1+  i\lambda_{i}}\right) f(z_{i},\bar{z}_{i})  \prod_{i=1}^{4}\textbf{1} _{[0,1]}(\s_i^\ast) 
\end{split}
\end{equation}
 
Then using the following identity\footnote{Note that there is no factor of $1/2$ here as opposed to equation \eqref{deltafuncid}. This is because the indicator function $\textbf{1} _{[0,1]}(\s_{1}^\ast)$ ensures that $\sigma_{1} \in [0,1]$.}
\begin{equation}
\label{deltafuncid1}
\begin{split}
& \delta (x) = \lim_{\epsilon \to 0 } \epsilon \hspace{0.04cm} x ^{\epsilon-1}, \quad\quad (0 \le x \le 1)
\end{split}
\end{equation}

we get,
\begin{equation}
\label{confsoftmhv1}
\begin{split}
 \lim_{\lambda_{1} \to 0} i \lambda_{1} \left[\lim_{\delta\rightarrow 0+}\frac{\Gamma (2+i\Lambda)}{\left(i \sum \varepsilon_i \sigma_i ^\ast u_i + \delta\right)^{2+i\Lambda}}\right] \sigma_{1}^{\ast i \lambda_{1}-1}& =  \left[\lim_{\delta\rightarrow 0+}\frac{\Gamma (2+i\Lambda')}{\left(i \sum \varepsilon_i \sigma_i ^\ast u_i + \delta\right)^{2+i\Lambda'}}\right] \delta(\sigma_{1}^{*})\\
\end{split}
\end{equation}

where $\Lambda'= \sum\limits_{i=2}^{4}\lambda_{i}$. Assuming $\bar{z}_{34}\ne 0$ the delta function in the above expression gives

\begin{equation}
\label{confsoftmhv2}
\begin{split}
\delta(\sigma_{1}^{*}) = \delta\left( -\frac{\varepsilon_{1}\varepsilon_{4}}{D}\frac{\bar{z}_{34}z_{24}}{\bar{z}_{13}z_{12}}\right)=\left| \frac{z_{12}\bar{z}_{13}}{\bar{z}_{34}} D\right|  \delta(z_{24})
\end{split}
\end{equation}

Further on the support of $\delta\left(z_{24}\right)$, we have

\begin{equation}
\label{confsoft2}
\begin{split}
 f(z_{i},\bar{z}_{i}) & \rightarrow \frac{z_{12}^{3}\bar{z}_{34}^{4}}{z_{13}z_{14}\bar{z}_{12}\bar{z}_{13}\bar{z}_{14}} \left|\frac{1}{z_{12}\bar{z}_{13}\bar{z}_{24}}\right| \delta\left( z_{34}\right)\\
 & = \frac{\bar{z}_{34}^{4}}{\bar{z}_{12}\bar{z}_{13}\bar{z}_{14}} \left | \frac{1}{\bar{z}_{13}\bar{z}_{24}}\right | \mathrm{sgn}(z_{12}) \delta(z_{34})
\end{split}
\end{equation}

Now due to the delta functions $\delta\left(z_{24}\right)$  and $\delta\left(z_{34}\right)$, the remaining simplex variables $\sigma^{*}_{2}, \sigma^{*}_{3}$ and $\sigma^{*}_{4}$  become

\begin{equation}
\label{confsoft3}
\begin{split}
& \sigma_{2}^{*} \rightarrow \rho^{*}_{2}= \varepsilon_{2}\varepsilon_{4}\frac{\bar{z}_{43}}{D_{3}}, \quad \sigma_{3}^{*}\rightarrow \rho^{*}_{3}= -\varepsilon_{3}\varepsilon_{4}\frac{\bar{z}_{42}}{D_{3}}, \quad \sigma_{4}^{*}\rightarrow \rho^{*}_{4}=\frac{\bar{z}_{32}}{ D_{3}} \\
\end{split}
\end{equation}

where
\begin{equation}
\label{dprimedef}
\begin{split}
D \rightarrow \frac{D_{3}}{\bar{z}_{23}}, \quad D_{3}= (1-\varepsilon_{4}\varepsilon_{3}) \bar{z}_{42}+  (\varepsilon_{4}\varepsilon_{2}-1)\bar{z}_{43}
\end{split}
\end{equation}



Combining the above results we then obtain

\begin{equation}
\label{confsoft4}
\begin{split}
 \left(\prod_{i=2}^{4} \sigma_{i}^{\ast 1+  i\lambda_{i}}\right) f(z_{i},\bar{z}_{i})  \delta(\sigma_{1}^{*}) & \rightarrow \left(\prod_{i=2}^{4} \rho_{i}^{\ast  i\lambda_{i}}\right) \frac{z_{12}}{\bar{z}_{12}} \frac{\bar{z}_{43}^{4}}{\bar{z}_{13}\bar{z}_{14}D_{3}^{2}} \hspace{0.05cm} \mathrm{sgn}(\varepsilon_{2} \varepsilon_{3} \bar{z}_{34}\bar{z}_{42}\bar{z}_{32}D_{3}) \delta(z_{24})\delta(z_{34}) \\
 &= \left(\prod_{i=2}^{4} \rho_{i}^{\ast  i\lambda_{i}}\right) \frac{z_{12}}{\bar{z}_{12}} \frac{\bar{z}_{43}^{4}}{\bar{z}_{13}\bar{z}_{14}D_{3}^{2}} \hspace{0.05cm}  \delta(z_{24})\delta(z_{34})
\end{split}
\end{equation}

Note that the last line in the above expression follows from the fact that due to the $3$-particle indicator function $\prod_{i=2}^{4}\textbf{1} _{[0,1]}(\sigma_i^\ast)$, we get $ \mathrm{sgn}(\varepsilon_{2} \varepsilon_{3} \bar{z}_{34}\bar{z}_{42}\bar{z}_{32}D_{3})=\mathrm{sgn}\left(\rho^{*}_{2}\rho^{*}_{3}\rho^{*}_{4}\right)=1$.\\

Now equation \eqref{confsoftmhv} takes the form

\begin{equation}
\label{confsoft5}
\begin{split}
& \lim_{\lambda_{1}\to 0} i \lambda_{1} \hspace{0.05cm} \mathcal{A}(1^{--},2^{--},3^{++},4^{++})(u_i,z_i,\bar{z}_i,\l_i)\\
& = \frac{\kappa^{2}}{4}  \left[\lim_{\delta\rightarrow 0+}\frac{\Gamma (2+i\Lambda')}{\left(i \sum \varepsilon_i \rho_i ^\ast u_i + \delta\right)^{2+i\Lambda'}}\right] \left(\prod_{i=2}^{4} \rho_{i}^{\ast  i\lambda_{i}}\right) \frac{z_{12}}{\bar{z}_{12}} \frac{\bar{z}_{43}^{4}}{\bar{z}_{13}\bar{z}_{14}D_{3}^{2}} \hspace{0.05cm}  \delta(z_{24})\delta(z_{34}) \prod_{i=2}^{4}\textbf{1} _{[0,1]}(\rho_i^\ast)\\
& = - \frac{ \kappa}{2} \hspace{0.04cm}\varepsilon_{3}\varepsilon_{4}  \left[\lim_{\delta\rightarrow 0+}\frac{1+i\Lambda'}{\left(i \sum \varepsilon_i \rho_i ^\ast u_i + \delta\right)}\right] \rho^{*}_{2}  \left( \frac{z_{12}}{\bar{z}_{12}}  \frac{\bar{z}_{32}\bar{z}_{42}}{\bar{z}_{31}\bar{z}_{41}}\right) \hspace{0.05cm}\mathcal{A}(2^{--},3^{++},4^{++})(u_i,z_i,\bar{z}_i,\l_i)
\end{split}
\end{equation}

where $\mathcal{A}(2^{--},3^{++},4^{++})(u_i,z_i,\bar{z}_i,\l_i)$ is the modified Mellin transform of the $3$-point $\overline{\mathrm{MHV}}$ graviton amplitude $S(2^{--},3^{++},4^{++})(\om_{i},z_{i},\bar{z}_{i})$.This was evaluated in section \eqref{antimhvgrav} and is given by

\begin{equation}
\label{mellin3grav}
\begin{split}
& \mathcal{A}(2^{--},3^{++},4^{++})(u_i,z_i,\bar{z}_i,\l_i)\\
& = \kappa \left[\lim_{\delta\rightarrow 0+}\frac{\Gamma (1+i\Lambda')}{\left(i \sum \varepsilon_i \rho_i ^\ast u_i + \delta\right)^{1+i\Lambda'}}\right] \left(\prod_{i=2}^{4} \rho_{i}^{\ast  i\lambda_{i}}\right) \left(\frac{\rho^{*}_{4}\rho^{*}_{3}}{\rho^{*}_{2}} \frac{\bar{z}_{43}^{3}}{\bar{z}_{32}\bar{z}_{24}} \right) ^{2} \frac{\delta(z_{24})\delta(z_{34})}{ 2 \rho^{*}_{4}\rho^{*}_{3}\rho^{*}_{2}D_{3}^{2}} \prod_{i=2}^{4}\textbf{1} _{[0,1]}(\s_i^\ast)
\end{split}
\end{equation}

Let us now differentiate the above amplitude with respect to $u_{2}$. This yields,

\begin{equation}
\label{mellin3grav1}
\begin{split}
& \frac{\partial}{\partial u_{2}}  \mathcal{A}(2^{--}3^{++}4^{++})= - \left[\lim_{\delta\rightarrow 0+}\frac{1+i\Lambda'}{\left(i \sum \varepsilon_i \rho_i ^\ast u_i + \delta\right)}\right] i \epsilon_{2} \rho^{*}_{2} \hspace{0.05cm}\mathcal{A}(2^{--},3^{++},4^{++})\\
\end{split}
\end{equation}

Using this in equation \eqref{confsoft5} we finally get
 
\begin{equation}
\label{confsoft6}
\begin{split}
& \lim_{\lambda_{1}\to 0} i \lambda_{1} \hspace{0.05cm} \mathcal{A}(1^{--},2^{--},3^{++},4^{++})(u_i,z_i,\bar{z}_i,\l_i)\\
&=  \frac{ \kappa}{2} \left(- i \varepsilon_{2}\varepsilon_{3}\varepsilon_{4}   \frac{z_{12}}{\bar{z}_{12}}  \frac{\bar{z}_{32}\bar{z}_{42}}{\bar{z}_{31}\bar{z}_{41}}\right) \frac{\partial}{\partial u_{2}} \left( \mathcal{A}(2^{--}3^{++}4^{++})(u_i,z_i,\bar{z}_i,\l_i) \right) \\
& =  \frac{ \kappa}{2}  \left(-i\varepsilon_{1}   \frac{z_{12}}{\bar{z}_{12}}  \frac{\bar{z}_{32}\bar{z}_{42}}{\bar{z}_{31}\bar{z}_{41}}\right) \frac{\partial}{\partial u_{2}} \left( \mathcal{A}(2^{--}3^{++}4^{++})(u_i,z_i,\bar{z}_i,\l_i) \right) 
\end{split}
\end{equation}

In going from the second to the last line in the above we have used the fact $\varepsilon_{1}\varepsilon_{2}\varepsilon_{3}\varepsilon_{4}=1$ when either two particles are incoming and two particles are outgoing or all $4$ particles are incoming (outgoing). 

Evidently the result in \eqref{confsoft6} agrees with the general case considered in equation \eqref{conformalsoftgrav3} if we choose the reference spinors to be $x=3, y=4$.

\subsection{Conformal soft limit in gluon four point amplitude}

Conformal soft limit of celestial amplitudes in gauge theory has been studied in \cite{Pate:2019mfs}.   In this section we show how to take soft limit in modified Mellin transformed gauge amplitude. The 4-point modified Mellin MHV amplitude in Yang-Mills theory in $\left(-,+,+,+\right)$ spacetime signature obtained in Eq.(\ref{gluon-4pt}) can be expressed as
\begin{eqnarray}\label{4-pt-gluon}
&&\mathcal{A}\left(1^{- }2^{-} 3^{+} 4^+\right))(u_{i} ,z_{i},\bar{z}_{i},\lambda_{i})\nonumber\\
&= &\frac{1}{4}\left[ \lim_{\delta\rightarrow 0+}\frac{\Gamma\left(i\sum\limits_{i=1}^{4} \lambda_{i}\right)}{\left(i\sum\varepsilon_{i} \sigma_{i}^{\ast} u_{i} +\delta  \right)^{i\sum\limits_{i=1}^{4} \lambda_{i}}} \right]\left(\prod\limits_{i=1}^{4} \sigma_{i}^{\ast i\lambda_{i}}\right)\frac{\sigma_{1}^{\ast} \sigma_{2}^{\ast}}{\sigma_{3}^{\ast} \sigma_{4}^{\ast}}\frac{z_{12}^3}{z_{23}z_{34}z_{41}}\frac{\delta(|z-\bar{z}|)}{ z_{13}\bar{z}_{13}z_{24}\bar{z}_{24}}\prod_{i=1}^{4}\textbf{1} _{[0,1]}(\sigma_{i}^{\ast}) \nonumber\\
& = & \frac{1}{4}\left[ \lim_{\delta\rightarrow 0+}\frac{\Gamma\left(i\Lambda\right)}{\left(i\varepsilon_{4}\bigl\{-\frac{z_{24}\bar{z}_{34}}{z_{12}\bar{z}_{13}}u_{1} + \frac{z_{34}\bar{z}_{14}}{z_{23}\bar{z}_{12}}u_{2} - \frac{z_{24}\bar{z}_{14}}{z_{23}\bar{z}_{12}}u_{3} + u_{4}\bigr\} +\delta  \right)^{i\Lambda}} \right] \left(\frac{\varepsilon_{3}}{\varepsilon_{1}}\frac{z_{23}\bar{z}_{34}}{z_{12}\bar{z}_{14}}\right)^{1+i\lambda_{1}}\left(\frac{\varepsilon_{3}}{\varepsilon_{2}}\frac{z_{13}\bar{z}_{34}}{z_{12}\bar{z}_{42}}\right)^{1+i\lambda_{2}} \nonumber\\
&& \times \left(\frac{\varepsilon_{3}}{\varepsilon_{4}}\frac{\bar{z}_{13}z_{23}}{\bar{z}_{14}z_{42}}\right)^{-1+i\lambda_{4} -i\Lambda} \delta\left(|z_{12}z_{34}\bar{z}_{13}\bar{z}_{24} - z_{13}z_{24}\bar{z}_{12}\bar{z}_{34}|\right)\frac{z_{12}^3}{z_{23}z_{34}z_{41}}\prod_{i=1}^{4}\textbf{1} _{[0,1]}(\sigma_{i}^{\ast}) . \nonumber\\
\end{eqnarray}
Here $\Lambda = \sum\limits_{i=1}^{4}\lambda_{i}$. The indicator functions signify that $\sigma_{i}^{\ast}$ should be positive otherwise the amplitude vanishes. For two-to-two scattering process, $ij \rightleftarrows kl$ with $\varepsilon_{i} = \varepsilon _{j} = - \varepsilon_{k} = -\varepsilon_{l}$ the indicator functions imply
\begin{eqnarray}
12 \rightleftarrows 34 & \Rightarrow & z_{1}> z_{3} > z_{2} > z_{4} \quad \text{or} \quad z_{1}< z_{3} < z_{2} < z_{4}, \nonumber\\
13 \rightleftarrows 24 & \Rightarrow & z_{1} > z_{2} > z_{3} > z_{4} \quad \text{or} \quad z_{1}< z_{2}< z_{3}< z_{4},  \nonumber\\
14 \rightleftarrows 23 & \Rightarrow & z_{1} > z_{3} >z_{4} >z_{2} \quad \text{or} \quad z_{1}<z_{3}<z_{4}<z_{2}.
\end{eqnarray}
Similar orderings exist for $\bar{z}_{ij}$. It can be checked that above conditions are consistent with replacing the product of indicator functions by step functions as 
\begin{equation}\label{theta}
\prod_{i=1}^{4}\textbf{1} _{[0,1]}(\sigma_{i}^{\ast}) = \Theta\left(\frac{\varepsilon_{3}}{\varepsilon_{1}}\frac{z_{23}\bar{z}_{34}}{z_{12}\bar{z}_{14}}\right)\Theta\left(\frac{\varepsilon_{3}}{\varepsilon_{2}}\frac{z_{13}\bar{z}_{34}}{z_{12}\bar{z}_{42}}\right)\Theta\left(\frac{\varepsilon_{3}}{\varepsilon_{4}}\frac{\bar{z}_{13}z_{23}}{\bar{z}_{14}z_{42}}\right).
\end{equation}
Including the factor $\left(\frac{\varepsilon_{3}}{\varepsilon_{4}}\frac{\bar{z}_{13}z_{23}}{\bar{z}_{14}z_{42}}\right)^{-i\Lambda}$ denominator of Eq.(\ref{4-pt-gluon}) becomes 
\begin{equation}\label{deno}
\lim_{\delta\rightarrow 0+}\left[i\varepsilon_{3}\left(\frac{z_{23}\bar{z}_{34}}{z_{12}\bar{z}_{14}}u_{1} + \frac{z_{13}\bar{z_{34}}}{z_{12}\bar{z}_{24}}u_{2} + \frac{\bar{z}_{13}}{\bar{z}_{12}}u_{3} + \frac{\bar{z}_{13}z_{23}}{\bar{z}_{14}z_{42}} u_{4}\right) + \delta\right]^{i\Lambda}.
\end{equation}
To take soft limit we multiply both sides of Eq.(\ref{4-pt-gluon}) by $\lim_{\lambda_{4}\rightarrow 0} i\lambda_{4}$. Following Eq.(\ref{deltafuncid}) we get 
\begin{equation}
\lim_{\lambda_{4}\rightarrow 0} i\lambda_{4}\left(\frac{\varepsilon_{3}}{\varepsilon_{4}}\frac{\bar{z}_{13}z_{23}}{\bar{z}_{14}z_{42}}\right)^{-1+i\lambda_{4} }= 2\delta \left(\frac{\varepsilon_{3}}{\varepsilon_{4}}\frac{\bar{z}_{13}z_{23}}{\bar{z}_{14}z_{42}}\right) = \text{sgn}\left(\bar{z}_{14}z_{42}z_{23}\right)2\frac{\bar{z}_{14}z_{42}}{z_{23}}\delta\left(\bar{z}_{13}\right).
\end{equation}
Using the above relation delta function appearing in Eq.(\ref{4-pt-gluon}) becomes
\begin{equation}
\delta\left(|z_{12}z_{34}\bar{z}_{13}\bar{z}_{24} - z_{13}z_{24}\bar{z}_{12}\bar{z}_{34}|\right) = \frac{\text{sgn}\left(z_{13}z_{24}\bar{z}_{34}\right)}{z_{13}z_{24}\bar{z}_{34}}\delta\left(\bar{z}_{12}\right).
\end{equation}
These two constraints also imply $\bar{z}_{12} = \bar{z}_{13}  = \bar{z}_{23} = 0$. Then in the soft limit Eq.(\ref{deno}) takes the form
\begin{equation}
\lim_{\delta\rightarrow 0+}\left[i\varepsilon_{3} z_{12}^{-1}\left(z_{23}u_{1} + z_{13}u_{2} + z_{12}u_{3}\right) + \delta\right]^{i\sum\limits_{i=1}^{3}\lambda_{i}}.
\end{equation}
We note that in the soft limit argument inside the last $\Theta$ function in Eq.(\ref{theta}) vanishes which gives $\Theta\left(0\right) = \frac{1}{2}$. Then  combining the delta functions and other $z$-dependent factors we get
\begin{eqnarray}
&&\lim_{\lambda_{4}\rightarrow 0} i \lambda_{4} \mathcal{A}\left(1^{- }2^{-} 3^{+} 4^+\right)(u_{i} ,z_{i},\bar{z}_{i},\lambda_{i})\nonumber\\
& = & \text{sgn}\left(z_{23}z_{31}\right)\frac{z_{31}}{z_{34}z_{41}} \delta\left(\bar{z}_{12}\right)\delta\left(\bar{z}_{13}\right)  \varepsilon_{1}\varepsilon_{2}\varepsilon_{1}^{i\lambda_{1}}\varepsilon_{2}^{i\lambda_{2}}\varepsilon_{3}^{i\lambda_{3}}z_{12}^{1+i\lambda_{3}}z_{23}^{-1+i\lambda_{1}}z_{31}^{-1+i\lambda_{2}} \nonumber\\
&& \frac{1}{4}\lim_{\delta\rightarrow 0}\frac{\Gamma\left(i\left(\lambda_{1} + \lambda_{2} + \lambda_{3}\right)\right)}{\left[i\left(u_{1}z_{23} + u_{2}z_{31} + u_{3}z_{12}\right) + \delta\right]^{i\left(\lambda_{1} + \lambda_{2} + \lambda_{3}\right)}} \Theta\left(\frac{\varepsilon_{3}}{\varepsilon_{1}}\frac{z_{23}}{z_{12}}\right)\Theta\left(\frac{\varepsilon_{3}}{\varepsilon_{2}}\frac{z_{31}}{z_{12}}\right) \nonumber\\
& = & -\frac{1}{2}\frac{z_{31}}{z_{34}z_{41}} \mathcal{A}\left(1^{-}2^{-}3^{+}\right).
\end{eqnarray}
Here the 3-point gluon amplitude is expressed in spacetime with split signature $\left(-,+,-,+\right)$. In this case 3-point gluon amplitude can be worked out in analogous way to the graviton 3-point amplitude.

\section*{Acknowledgement}
We would like to thank Alok Laddha and Ashoke Sen for valuable discussions.  This research was supported in part by the International Centre for Theoretical Sciences (ICTS) during a visit for participating in the program - AdS/CFT at 20 and Beyond (Code: ICTS/adscft20/05). SG would like to thank Yasha Neiman, Vyacheslav Lysov and Lin-Qing Chen for useful discussions and acknowledges the hospitality of the Korea Institute for Advanced Study (KIAS), Seoul, South Korea while this work was being completed. SG is supported by the Quantum Gravity Unit of the Okinawa Institute of Science and Technology Graduate University (OIST). 



\end{document}